# Reify Your Collection Queries for Modularity and Speed!

## Extended Version

Paolo G. Giarrusso
Philipps University Marburg

Klaus Ostermann
Philipps University Marburg

Michael Eichberg
Software Technology Group,
Technische Universität
Darmstadt

Ralf Mitschke
Software Technology Group,
Technische Universität
Darmstadt

Tillmann Rendel
Philipps University Marburg

Christian Kästner
Carnegie Mellon University

## ABSTRACT

Modularity and efficiency are often contradicting requirements, such that programers have to trade one for the other. We analyze this dilemma in the context of programs operating on collections. Performance-critical code using collections need often to be hand-optimized, leading to non-modular, brittle, and redundant code. In principle, this dilemma could be avoided by automatic collection-specific optimizations, such as fusion of collection traversals, usage of indexing, or reordering of filters. Unfortunately, it is not obvious how to encode such optimizations in terms of ordinary collection APIs, because the program operating on the collections is not reified and hence cannot be analyzed.

We propose SQUOPT, the Scala Query Optimizer—a *deep embedding* of the Scala collections API that allows such analyses and optimizations to be defined and executed within Scala, without relying on external tools or compiler extensions. SQUOPT provides the same "look and feel" (syntax and static typing guarantees) as the standard collections API. We evaluate SQUOPT by re-implementing several code analyses of the Findbugs tool using SQUOPT, show average speedups of 12x with a maximum of 12800x and hence demonstrate that SQUOPT can reconcile modularity and efficiency in real-world applications.

## 1. INTRODUCTION

In-memory collections of data often need efficient processing. For on-disk data, efficient processing is already provided by database management systems (DBMS), thanks to their query optimizers which support many optimizations specific to the domain of collections; moving in-memory data to DBMSs however does not typically improve performance [39], and query optimizers cannot be reused separately since DBMS are typically monolithic with deeply integrated optimizers. A few collection-specific optimizations, such as shortcut fusion [12] are supported by compilers for purely functional languages such as Haskell, but the implementation techniques do not generalize to many other optimizations such as support for indexes. In general, collection-specific optimizations are not supported by the general-purpose optimizers used by typical (JIT) compilers.

Therefore, when collection-related optimizations are needed, programmers perform them by hand. Some optimizations are not hard to apply manually, but in many cases become applicable only after manual inlining [32]. But manual inlining modifies source code by combining distinct functions together, while often distinct functions should remain distinct because they deal with different concerns, or because one of the them is reused in other contexts. In both cases, manual inlining reduces modularity.

For these reasons, currently developers need to choose between modularity and performance, as also highlighted by Kiczales et al. [21]. Instead, we envision that they should rely on an automatic optimizer performing inlining and collection-specific optimizations. Then they would achieve both performance and modularity.

One way to implement such an optimizer would be to extend the compiler of the language with a collection-specific optimizer, or to add some kind of external preprocessor to the language. However, such solutions would be rather brittle (for instance, they lack composability with other language extensions) and they would preclude optimization opportunities that arise only at runtime.

For this reason, our approach is implemented as an embedded domain-specific language, that is, as a regular library. We call this library SQUOPT, the Scala QUery OPTimizer. SQUOPT consists of a domain-specific language (DSL) for queries on collections based on the Scala collections API. This DSL is implemented as an embedded DSL (EDSL) for Scala. An expression in this EDSL produces at run time an *expression tree* in the host language: a data structure which represents the query to execute, similar to an abstract syntax tree (AST). Thanks to the extensibility of Scala, expressions in this language look almost identical to expressions with the same meaning in Scala. Again at run time, SQuOpt optimizes and compiles these expression trees for more efficient execution. Doing optimization at run time, instead of compile-time, obviates the need for control-flow analyses to determine which code will be actually executed [3], as we will see later.

We have choosen Scala [30] to implement our library for two reasons: i) Scala is a good meta-language for embedded DSLs, because it is syntactically flexible and has a powerful type system, and ii) Scala has a sophisticated collections library with an attractive syntax (for-comprehensions) to specify queries.

To evaluate SQuOpt, we study queries of the Findbugs tool [20]. We rewrote a set of queries to use the Scala collections API and show that modularization incurs significant performance overhead. Subsequently, we consider versions of the same queries using SQuOpt. We demonstrate that the automatic optimization can reconcile modularity and performance in many cases. Adding advanced optimizations such as indexing can even improve the performance of the analyses beyond the original non-modular analyses.

Overall, our main contributions are the following:

- We illustrate the tradeoff between modularity and performance when manipulating collections, caused by the lack of domain-specific optimizations (Sec. 2). Conversely, we illustrate how domain-specific optimizations lead to more readable and more modular code (Sec. 3).

- We present the design and implementation of SQuOpt, an embedded DSL for queries on collections in Scala (Sec. 4).

- We evaluate SQuOpt to show that it supports writing queries that are at the same time modular and fast. We do so by re-implementing several code analyses of the Findbugs tool. The resulting code is more modular and/or more efficient, in some cases by orders of magnitude. In these case studies, we measured average speedups of 12x with a maximum of 12800x (Sec. 5).

## 2. MOTIVATION

In this section, we show how the absense of collection-specific optimizations forces programmers to trade modularity against performance, which motivates our design of SQuOpt to resolve this conflict.

As our running example through the paper, we consider representing and querying a simple in-memory bibliography. A book has, in our schema, a title, a publisher and a list of authors. Each author, in turn, has a first and last name. We represent authors and books as instances of the Scala classes `Author` and `Book` shown in Fig. 1. The class declarations list the type of each field: Titles, publishers, and first and last names are all stored in fields of type `String`. The list of authors is stored in a field of type `Seq[Author]`, that is, a sequence of authors. The code fragment also defines a collection of books named `books`.

As a common idiom to query such collections, Scala provides *for-comprehensions*. For instance, the for-comprehension in Fig. 2 finds all books published by Pearson

```
package schema
case class Author(firstName: String, lastName: String)
case class Book(title: String, publisher: String,
  authors: Seq[Author])

val books: Set[Book] = Set(
  new Book("Compilers: Principles, Techniques and Tools",
      "Pearson Education",
      Seq(new Author("Alfred V.", "Aho"),
          new Author("Monica S.", "Lam"),
          new Author("Ravi", "Sethi"),
          new Author("Jeffrey D.", "Ullman")))
  /* other books ... */)
```

Figure 1: Definition of the schema and of some content.

```
case class BookData(title: String, authorName: String,
  coauthors: Int)

val records =
  for {
    book ← books
    if book.publisher == "Pearson Education"
    author ← book.authors
  } yield new BookData(book.title,
              author.firstName + " " +
              author.lastName,
              book.authors.size - 1)
```

Figure 2: Our example query on the schema in Fig. 1.

Education and yields, for each of those books, and for each of its authors, a record containing the book title, the full name of that author and the number of additional coauthors. The statement `book ← books` functions like a loop header: The remainder of the for-comprehension is executed once per book in the collection. Consequently, the statement `author ← book.authors` starts a nested loop. The return value of the for-comprehension is a collection of all yielded records. Note that if a book has multiple authors, this for-comprehensions will return multiple records relative to this book, one for each author.

We can further process this collection with another for-comprehension, possibly in a different module. For example, the function in Fig. 3 filters book titles containing the word "Principles", and drops the number of coauthors from the result.

In Scala, the implementation of for-comprehensions is not fixed. Instead, the compiler desugars a for-comprehension to a series of API calls, and different collection classes can implement this API differently. Later, we will use this flexibility to provide an optimizing implementation of for-comprehensions, but in this section, we focus on the behavior of the standard Scala collections, which implement for-comprehensions as loops that create intermediate collections.

### 2.1 Optimizing by hand

The naive implementation in Fig. 2 and Fig. 3 is modular but inefficient:

1. To execute this code, we first build the original collection and only later we perform further processing to

```
def titleFilter(records: Set[BookData],
    keyword: String) =
  for {
    record ← records
    if record.title.contains(keyword)
  } yield (record.title, record.authorName)

val res = titleFilter(records, "Principles")
```

Figure 3: Another query, which processes the results of the query in Fig. 2.

   build the new result; creating the intermediate collection is costly.

2. The same book can appear in records more than once if the book has more than one author, but all of these duplicates have the same title. Nevertheless, we test each duplicate title separately whether it contains the searched keyword. If books have 4 authors on average, this means a slowdown of a factor of 4 for the filtering step.

The existing Scala collections API offers some generic concepts for optimization, such as non-strict variants of the query operators (called 'views' in Scala), but they can only be used for a limited set of optimizations, as we discuss in the section on related work. In general, one can only resolve these inefficiencies by manually optimizing the query; however, we will observe that the resulting code is less modular.

To address the first problem above, we can manually inline titleFilter and records; we obtain two nested for-comprehensions. Furthermore, we can *unnest* the inner one [7].

To address the second problem above, we *hoist* the filtering step, that is, we change the order of the processing steps in the query to first look for keyword within book.title and then iterate over the set of authors. This does not change the overall semantics of the query because the filter only accesses the title but does not depend on the author. In the end, we obtain the code in Fig. 4. The resulting query processes the title of each book only once. Since filtering in Scala is done lazily, the resulting query avoids building an intermediate collection.

This second optimization is only possible after inlining and thereby reducing the modularity of the code, because it mixes together processing steps from titleFilter and from the definition of records.

To make titleFilterHandOpt more reusable, we could turn the publisher name into a parameter. However, the new versions of titleFilter cannot be reused as-is if some details of the inlined code change; for instance, we might need to filter publishers differently or not at all. On the other hand, if we express queries modularly, we might lose some opportunities for optimization. The design of the Scala collections API forces us to manually optimize our code by repeated inlining and subsequent application of query optimization rules, which leads to a loss of modularity.

## 3. AUTOMATIC OPTIMIZATION WITH SQUOPT

The goal of SQuOpt is to let programmers write queries modularly and on a high-level of abstraction and deal with

```
def titleFilterHandOpt(books: Set[Book],
                       publisher: String,
                       keyword: String) =
  for {
    book ← books
    if book.publisher == publisher &&
       book.title.contains(keyword)
    author ← book.authors
  } yield (book.title, author.firstName + " " +
           author.lastName)
val res = titleFilterHandOpt(books,
  "Pearson Education", "Principles")
```

Figure 4: Composition of queries in Fig. 2 and Fig. 3, after inlining, query unnesting and hoisting.

```
import squopt._
import schema.squopt._

val recordsQuery =
  for {
    book ← books.asSquopt
    if book.publisher ==# "Pearson Education"
    author ← book.authors
  } yield new BookData(book.title,
    author.firstName + " " + author.lastName,
    book.authors.size - 1)

// ...
val records = recordsQuery.eval
```

Figure 5: SQuOpt version of Fig. 2; `recordQuery` contains a reification of the query, `records` its result.

optimization by a dedicated domain-specific optimizer. In our concrete example, programmers should be able to write queries similar to Fig. 2 and Fig. 3, but get the efficiency of Fig. 4. SQuOpt achieves this by overloading the implementation of for-comprehensions as well as the implementation of operations such as string concatenation with + and field access book.author in order to reify the query as an expression tree, optimize this expression tree, and then execute the optimized query. Programmers explicitly trigger SQuOpt by adapting their queries.

### 3.1 Adapting a Query

With SQuOpt, the first part of the running example becomes what is shown in Fig. 5. To use SQuOpt instead of native Scala queries, we first assume that the query is already purely functional. We argue that purely functional queries are more declarative. A main reason for using side effects is to improve performance, but SQuOpt voids this reason by automatically removing performance overhead by optimization. At the same time, the lack of side effects makes more optimizations possible.

Once the query is purely functional, a programmer needs to (a) import the SQuOpt library, (b) import some wrapper code specific to the types the collection operates on, in this case Book and Author (more about that later), (c) convert explicitly the native Scala collections involved to collections of our framework by a call to asSquopt, (d) rename a few operators such as == to ==# (this is necessary due to some Scala limitations), and (e) add a separate step where the query is evaluated (possibly after optimization). All these changes are lightweight and mostly of a syntactic nature.

```
def titleFilterQuery(records: Exp[Set[BookData]],
                     keyword: Exp[String]) = for {
  record ← records
  if record.title.contains(keyword)
} yield (record.title, record.authorName)
val resQuery = titleFilterQuery(recordsQuery, "Principles")
val res = resQuery.optimize.eval
```

Figure 6: SQuOpt version of Fig. 3

Fig. 6 shows how the second part of our running example can be adapted to SQuOpt. The type annotations in the function reveal some details of our implementation: Expressions that are reified have type Exp[T] instead of T. As the code shows, resQuery is optimized before compilation. This call will perform the optimizations we did by hand and will return a query equivalent to Fig. 4, after verifying their safety conditions. For instance, after inlining the filter `if book.title.contains(keyword)` does not reference author, hence it is safe to hoist it. Note that checking this safety condition would not be possible without a reification of the predicate. For instance, it would not be sufficient to only reify the calls to the collection API, because the predicate is represented as a boolean function parameter. In general, our automatic optimizer inspects the whole reification of the query implementation to check that optimizations do not introduce changes in the overall result of the query and are therefore safe.

### 3.2 Indexing

SQuOpt also supports the transparent usage of indexes. Indexes can further improve the efficiency of queries, sometimes by orders of magnitude. In our running example, the query scans all books to look for the ones having the right publisher. To speed up this query, we can preprocess books to build an index, that is, a dictionary mapping, from each publisher to a collection of all the books it published. This index can then be used to answer the original query directly without scanning all the books again.

We construct a *query* representing the desired dictionary, and inform the optimizer that it should use this index where appropriate:

```
val idxByPublisher =
  books.asSquopt.indexBy(_.publisher)
Optimization.addIndex(idxByPublisher)
```

The indexBy collection method accepts a function which maps a collection element to a key; `coll.indexBy(key)` returns a dictionary mapping each key to the collection of all elements of coll having that key. Missing keys are mapped to an empty collection.[1] Optimization.addIndex simply pre-evaluates the index and updates a dictionary mapping the index to its pre-evaluated result.

A call to optimize on a query will then take this index into account and rewrite the query to perform index lookup instead of scanning, if possible. For instance, the code in Fig. 5 would be rewritten by the optimizer to an output similar to the following query:

```
val indexedQuery =
```

---
[1] For readers familiar with the Scala collection API, we remark that the only difference with the standard groupBy method is the handling of missing keys.

```
  for {
    book ← idxByPublisher("Pearson Education")
    author ← book.authors
  } yield new BookData(book.title, author.firstName
      + " " + author.lastName, book.authors.size - 1)
```

Since dictionaries in Scala are functions, in the above code, dictionary lookup on idxByPublisher is represented simply as function application. The above code iterates over books having the desired publisher, instead of scanning the whole library, and performs the remaining computational steps from the original query. Although the index usage in the listing above is notated as `idxByPublisher("Pearson Education")`, only the cached result of evaluating the index is used when the query is executed, not the reified index definition.

This optimization could also be performed manually, of course, but the queries are on a higher abstraction level and more maintainable if indexing is defined separately and applied automatically. Manual application of indexing is a cross-cutting concern because adding or removing an index affects potentially many queries. SQuOpt does not free the developer from the task of assessing which index will 'pay off' (we have not considered automatic index creation), but at least it becomes simple to add or remove an index, since the application of the indexes is modularized in the optimizer.

## 4. IMPLEMENTATION

After describing how to use SQuOpt we explain how SQuOpt represents queries internally and optimizes them. Here we give only a brief overview of our implementation technique; it is described in more detail in Appendix B.

### 4.1 Expression Trees

In order to analyze and optimize collection queries at runtime, SQuOpt reifies their syntactic structure as *expression trees*. The expression tree reflects the syntax of the query after desugaring, that is, after for-comprehensions have been replaced by API calls. For instance, recordsQuery from Fig. 5 points to the following expression tree (with some boilerplate omitted for clarity):

```
new FlatMap(
  new Filter(
    new Const(books),
    v2 ⇒ new Eq(new Book_publisher(v2),
                new Const("Pearson Education"))),
    v3 ⇒ new MapNode(
          new Book_authors(v3),
          v4 ⇒ new BookData(
                new Book_title(v3),
                new StringConcat(
                  new StringConcat(
                    new Author_firstName(v4),
                    new Const(" ")),
                  new Author_lastName(v4)),
                new Plus(new Size(new Book_authors(v3)),
                         new Negate(new Const(1)))))))
```

The structure of the for-comprehension is encoded with the FlatMap, Filter and MapNode instances. These classes correspond to the API methods that for-comprehensions get desugared to. SQuOpt arranges for the implementation of flatMap to construct a FlatMap instance, etc. The instances of the other classes encode the rest of the structure of the collection query, that is, which methods are called on which arguments. On the one hand, SQuOpt defines classes such as Const or Eq that are generic and applicable to all queries.

On the other hand, classes such as `Book_publisher` cannot be predefined, because they are specific to the user-defined types used in a query. SQUOPT provides a small code generator, which creates a case class for each method and field of a user-defined type. Functions in the query are represented by functions that create expression trees; representing functions in this way is frequently called higher-order abstract syntax [34].

We can see that the reification of this code corresponds closely to an abstract syntax tree for the code which is executed; however, many calls to specific methods, like `map`, are represented by special nodes, like `MapNode`, rather than as method calls. For the optimizer it becomes easier to match and transform those nodes than with a generic abstract syntax tree.

Nodes for collection operations are carefully defined by hand to provide them highly generic type signatures and make them reusable for all collection types. In Scala, collection operations are highly polymorphic; for instance, `map` has a single implementation working on all collection types, like `List`, `Set`, and we similarly want to represent all usages of `map` through instances of a single node type, namely `MapNode`. Having separate nodes `ListMapNode`, `SetMapNode` and so on would be inconvenient, for instance when writing the optimizer. However, `map` on a `List[Int]` will produce another `List`, while on a `Set` it will produce another `Set`, and so on for each specific collection type (in first approximation); moreover, this is guaranteed statically by the type of `map`. Yet, thanks to advanced typesystem features, `map` is defined only once avoiding redundancy, but has a type polymorphic enough to guarantee statically that the correct return value is produced. Since our tree representation is strongly typed, we need to have a similar level of polymorphism in `MapNode`. We achieved this by extending the techniques described by Odersky and Moors [29], but cannot provide details in this context.

We get these expression trees by using Scala implicit conversions in a particular style, which we adopted from Rompf and Odersky [36]. Implicit conversions allow to add, for each method `A.foo(B)`, an overload of `Exp[A].foo(Exp[B])`. Where a value of type `Exp[T]` is expected, a value of type `T` can be used thanks to other implicit conversions, which wrap it in a `Const` node. The initial call of `asSquopt` triggers the application of the implicit conversions by converting the collection to the leaf of an expression tree.

It is also possible to call methods that do not return expression trees; however, such method calls would then only be represented by an opaque `MethodCall` node in the expression tree, which means that the code of the method cannot be considered in optimizations.

Crucially, these expression trees are generated at runtime. For instance, the first `Const` contains a reference to the actual collection of books to which `books` refers. If a query uses another query, such as `records` in Fig. 6, then the subquery is effectively *inlined*. The same holds for method calls inside queries: If these methods return an expression tree (such as the `titleFilterQuery` method in Fig. 6), then these expression trees are inlined into the composite query. Since the reification happens at runtime, it is not necessary to predict the targets of dynamically bound method calls: A new (and possibly different) expression tree is created each time a block of code containing queries is executed.

Hence, we can say that expression trees represent the computation which is going to be executed after inlining; control flow or virtual calls in the original code typically disappear—especially if they manipulate the query as a whole. This is typical of deeply embedded DSLs, where code instead of performing computations produces a representation of the computation to perform [5, 3].

This inlining can duplicate computations; for instance, after executing this code:

```
val num: Exp[Int] = 10
val square = num * num
val sum = square + square
```

evaluating `sum` will evaluate `square` twice. For this reason Elliott et al. [5] recommend to implement common-subexpression elimination, as we do.

## 4.2 Optimizations

Our optimizer currently supports several algebraic optimizations. Any query and in fact every reified expression can be optimized by calling the `optimize` function on it. The ability to optimize reified expressions that are not queries is useful; for instance, optimizing a function that produces a query is similar to a "prepared statement" in relational databases.

The optimizations we implemented are mostly standard in compilers [26] or databases:

- *Query unnesting* merges a nested query into the containing one [7, 16], replacing for instance

  **for** {val1 ← (**for** {val2 ← coll} **yield** f(val2))}
    **yield** g(val1)

  with

  **for** {val2 ← coll; val1 = f(val1)} **yield** g(val1)

- *Bulk operation fusion* fuses higher-order operators on collections.

- *Filter hoisting* tries to apply filters as early as possible; in database query optimization, it is known as selection pushdown . For filter hoisting, it is important that the full query is reified, because otherwise the dependencies of the filter condition cannot be determined.

- We reduce during optimization tuple/case class accesses: For instance, `(a, b)._1` is simplified to `a`. This is important because the produced expression does not depend on `b`; removing this false dependency can allow, for instance, a filter containing this expression to be hoisted to a context where `b` is not bound.

- *Indexing* tries to apply one or more of the available indexes to speed up the query.

- *Constant subexpression elimination (CSE)* avoids that the same computation is performed multiple times; we use techniques similar to Rompf and Odersky [36].

- Smaller optimizations include constant folding, reassociation of associative operators and removal of identity maps (`coll.map(x ⇒ x)`, typically generated by the translation of for-comprehensions).

Each optimization is applied recursively bottom-up until it does not trigger anymore; different optimizations are composed in a fixed pipeline.

Optimizations are only guaranteed to be semantics-preserving if queries obey the restrictions we mentioned: for instance, queries should not involve side-effects such as assignments or I/O, and all collections used in queries should implement the specifications stated in the collections API. Obviously the choice of optimizations involves many trade-offs; for that reason we believe that it is all the more important that the optimizer is not hard-wired into the compiler but implemented as a library, with potentially many different implementations.

To make changes to the optimizer more practical, we designed our query representation so that optimizations are easy to express; restricting to pure queries also helps. For instance, filter fusion can be implemented simply as: [2]

```
val mergeFilters = ExpTransformer {
  case Sym(Filter(Sym(Filter(collection, pred2)), pred1)) ⇒
    collection.filter(x ⇒ pred2(x) && pred1(x))
}
```

The above code matches on reified expression of form `collection.filter(pred2).filter(pred1)` and rewrites it. A more complex optimization like filter hoisting requires only 20 lines of code.

We have implemented a prototype of the optimizer with the mentioned optimizations. Many additional algebraic optimizations can be added in future work by us or others; a candidate would be loop hoisting, which moves out of loops arbitrary computations not depending on the loop variable (and not just filters). With some changes to the optimizer's architecture, it would also be possible to perform cost-based and dynamic optimizations.

### 4.3 Query execution

Calling the `eval` method on a query will convert it to executable bytecode; this bytecode will be loaded and invoked by using Java reflection. We produce a thunk that, when evaluated, will execute the generated code.

In our prototype we produce bytecode by converting expression trees to Scala code and invoking on the result the Scala compiler, `scalac`. Invoking `scalac` is typically quite slow, but it is merely an engineering problem to produce bytecode in a faster way; we currently use caching to limit this concern.

Our expression trees contain native Scala values wrapped in `Const` nodes, and in many cases one cannot produce Scala program text evaluating to the same value. To allow executing such expression trees we need to implement cross-stage persistence (CSP): the generated code will be a function, accepting the actual values as arguments [36]. This allows sharing the compiled code for expressions which differ only in the embedded values.

More in detail, our compilation algorithm is as follows. (a) We implement CSP by replacing embedded Scala values by references to the function arguments; so for instance `List(1, 2, 3).map(x ⇒ x + 1)` becomes the function `(s1: List[Int], s2: Int) ⇒ s1.map(x ⇒ x + s2)`. (b) We look up the produced expression tree, together with the types of the constants we just removed, in a cache mapping to the generated classes. If the lookup fails we update the cache with the result of the next steps. (c) We apply CSE on the expression. (d) We convert the tree to code, compile it and load the generated code.

---
[2] `Sym` nodes are part of the boilerplate we omitted earlier.

**Preventing errors in generated code** Compiler errors in generated code are typically a concern; with SQuOpt, however, they can only arise due to implementation bugs in SQuOpt (for instance in pretty-printing, which cannot be checked statically), so they do not concern users. Since our query language and tree representation are statically typed, type-incorrect queries will be rejected statically. For instance, consider again `idxByPublisher`, described previously:

```
val idxByPublisher =
  books.asSquopt.indexBy(_.publisher)
```

Since `Book.publisher` returns a `String`, `idxByPublisher` has type `Exp[Map[String, Book]]`. Looking up a key of the wrong type, for instance by writing `idxByPublisher(book)` where `book: Book`, will make `scalac` emit a static type error.

## 5. EVALUATION

The key goals of SQuOpt are to reconcile *modularity* and *efficiency*. To evaluate this claim, we perform a rigorous performance evaluation of queries with and without SQuOpt. We also analyze modularization potential of these queries and evaluate how modularization affects performance (with and without SQuOpt).

We show that modularization introduces a significant slowdown. The overhead of using SQuOpt is usually moderate, and optimizations can compensate this overhead, remove the modularization slowdown and improve performance of some queries by orders of magnitude, especially when indexes are used.

### 5.1 Study Setup

Throughout the paper, we have already shown several compact queries for which our optimizations increase performance significantly compared to a naive execution. Since some optimizations change the complexity class of the query (e.g. by using an index), so the speedups grow with the size of the data. However, to get a more realistic evaluation of SQuOpt, we decided to perform an experiment with existing real-world queries.

As we are interested in both performance and modularization, we have a specification and three different implementations of each query that we need to compare:

(0) **Query specification:** We selected a set of existing real-world queries specified and implemented independently from our work and prior to it. We used only the specification of these queries.

(1) **Modularized Scala implementation:** We reimplemented each query as an expression on Scala collections—our baseline implementation. For modularity, we separated reusable domain abstractions into subqueries. We confirmed the abstractions with a domain expert and will later illustrate them to emphasize their general nature.

(2) **Hand-optimized Scala implementation:** Next, we asked a domain expert to performed manual optimizations on the modularized queries. The expert should perform optimizations, such as inlining and filter hoisting, where he could find performance improvements.

(3) **SQuOpt implementation:** Finally, we rewrote the modularized Scala queries from (1) as SQuOpt queries. The rewrites are of purely syntactic nature to use our library (as described in Sec. 3.1) and preserve the modularity of the queries.

Since SQuOpt supports executing queries with and without optimizations and indexes, we measured actually three different execution modes of the SQuOpt implementation:

($3^-$) **SQuOpt without optimizer:** First, we execute the SQuOpt queries without performing optimization first, which should show the SQuOpt overhead compared to the modular Scala implementation (1). However, common-subexpression elimination is still used here, since it is part of the compilation pipeline. This is appropriate to counter the effects of excessive inlining due to using a deep embedding, as explained in Sec. 4.1.

($3^o$) **SQuOpt with optimizer:** Next, we execute the SQuOpt queries after optimization.

($3^x$) **SQuOpt with optimizer and indexes:** Finally, we execute the queries after providing a set of indexes that the optimizer can consider.

In all cases, we measure query execution time for the generated code, excluding compilation: we consider this appropriate because the results of compilations are cached aggressively and can be reused when the underlying data is changed, as the data is not part of the compiled code.

We use additional indexes in ($3^x$), but not in the hand-optimized Scala implementation (2). We argue that indexes are less likely to be applied manually, because index application is a crosscutting concern and makes the whole query implementation more complicated and less abstract. Still, we offer measurement ($3^o$) to compare the speedup without additional indexes.

This gives us a total of five settings to measure and compare (1, 2, $3^-$, $3^o$, and $3^x$). Between them, we want to observe the following interesting performance ratios (speedups or slowdowns):

(**M**) Modularization overhead (the relative performance difference between the modularized and the hand-optimized Scala implementation: $1/2$).

(**S**) SQuOpt overhead (the overhead of executing unoptimized SQuOpt queries: $1/3^-$; smaller is better).

(**H**) Hand-optimization challenge (the performance overhead of our optimizer against hand-optimizations of a domain expert: $2/3^o$; bigger is better). This overhead is partly due to the SQuOpt overhead (S) and partly to optimizations which have not been automated or have not been effective enough. This comparison excludes the effects of indexing, since this is an optimization we did not perform by hand; we also report (**H'**) = $2/3^x$, which includes indexing.

(**O**) Optimization potential (the speedup by optimizing modularized queries: $1/3^o$; bigger is better).

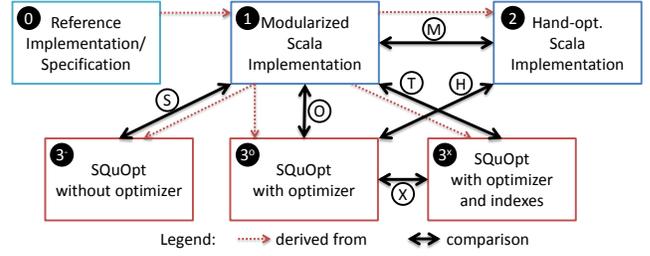

Figure 7: Measurement Setup: Overview

| Abstraction | Used |
|---|---|
| All fields in all class files | 4 |
| All methods in all class files | 3 |
| All method bodies in all class files | 3 |
| All instructions in all method bodies and their bytecode index | 5 |
| Sliding window (size $n$) over all instructions (and their index) | 3 |

Table 1: Description of abstractions removed during hand-optimization and number of queries where the abstraction is used (and optimized away).

(**X**) Index influence (the speedup gained by using indexes: $3^o/3^x$) (bigger is better).

(**T**) Total optimization potential with indexes ($1/3^x$; bigger is better), which is equal to $(O) \times (X)$.

In Figure 7, we provide an overview of the setup. We made our raw data available and our results reproducible [40].[3]

## 5.2 Experimental Units

As experimental units, we sampled a set of queries on code structures from Findbugs 2.0 [20]. Findbugs is a popular bug-finding tool for Java Bytecode available as open source. To detect bug patterns, Findbugs performs queries that traverse a structural in-memory representation of a code base (extracted from bytecode). We selected queries from Findbugs because they represent typical non-trivial queries on in-memory collections.

We sampled queries in two batches. First, we manually selected 8 queries (from approx. 400 queries in Findbugs), chosen mainly to evaluate the potential speedups of indexing (queries that primarily looked for declarations of classes, methods, or fields with specific properties, queries that inspect the type hierarchy, and queries that required analyzing methods implementation). Subsequently, we *randomly* selected a batch of 11 additional queries. The batch excluded queries that rely on control-/dataflow analyses (i.e., analyzing the effect of bytecode instructions on the stack), due to limitations of the bytecode tookit we use. In total, we have 19 queries as listed in Table 2 (the randomly selected queries are marked with [R]).

We implemented each query three times (see implementations (1)–(3) in Sec. 5.1) following the specifications given in the Findbugs documentation (0). Instead of using a hierarchy of visitors as the original implementations of the queries

---
[3]Data available at: http://www.informatik.uni-marburg.de/~pgiarrusso/SQuOpt

|     |                                              | Performance (ms) |      |       |       |       | Performance ratios |         |           |
| --- | -------------------------------------------- | ---- | ---- | ----- | ----- | ----- | ------- | -------- | --------- |
| Id  | Description                                  | 1    | 2    | $3^-$ | $3^o$ | $3^x$ | M (1/2) | H (2/$3^o$) | T (1/$3^x$) |
| 1   | Covariant compareTo() defined                | 1.1  | 1.3  | 0.85  | 0.26  | 0.26  | 0.9     | 5.0      | 4.4       |
| 2   | Explicit garbage collection call             | 496  | 258  | 1176  | 1150  | 52    | 1.9     | 0.2      | 9.5       |
| 3   | Protected field in final class               | 11   | 1.1  | 11    | 1.2   | 1.2   | 10.0    | 1.0      | 9.8       |
| 4   | Explicit runFinalizersOnExit() call          | 509  | 262  | 1150  | 1123  | 10.0  | 1.9     | 0.2      | 51        |
| 5   | clone() defined in non-Cloneable class       | 29   | 14   | 55    | 46    | 0.47  | 2.1     | 0.3      | 61        |
| 6   | Covariant equals() defined                   | 29   | 15   | 23    | 9.7   | 0.20  | 1.9     | 1.6      | 147       |
| 7   | Public finalizer defined                     | 29   | 12   | 28    | 8.0   | 0.03  | 2.3     | 1.5      | 1070      |
| 8   | Dubious catching of IllegalMonitorStateException | 82 | 72 | 110 | 28 | 0.01 | 1.1 | 2.6 | 12800 |
| $9^R$  | Uninit. field read during construction of super | 896 | 367 | 3017 | 960 | 960 | 2.4 | 0.4 | 0.9 |
| $10^R$ | Mutable static field declared public         | 9527 | 9511 | 9115 | 9350 | 9350 | 1.0 | 1.0 | 1.0 |
| $11^R$ | Refactor anon. inner class to static         | 8804 | 8767 | 8718 | 8700 | 8700 | 1.0 | 1.0 | 1.0 |
| $12^R$ | Inefficient use of toArray(Object[])         | 3714 | 1905 | 4046 | 3414 | 3414 | 2.0 | 0.6 | 1.1 |
| $13^R$ | Primitive boxed and unboxed for coercion     | 3905 | 1672 | 5044 | 3224 | 3224 | 2.3 | 0.5 | 1.2 |
| $14^R$ | Double precision conversion from 32 bit      | 3887 | 1796 | 5289 | 3010 | 3010 | 2.2 | 0.6 | 1.3 |
| $15^R$ | Privileged method used outside doPrivileged  | 505  | 302  | 1319  | 337   | 337   | 1.7     | 0.9      | 1.5       |
| $16^R$ | Mutable public static field should be final  | 13   | 6.2  | 12    | 7.0   | 7.0   | 2.0     | 0.9      | 1.8       |
| $17^R$ | Serializable class is member of non-ser. class | 12 | 0.77 | 0.94 | 1.8 | 1.8 | 16 | 0.4 | 6.9 |
| $18^R$ | Swing methods used outside Swing thread      | 577  | 53   | 1163  | 45    | 45    | 11      | 1.2      | 13        |
| $19^R$ | Finalizer only calls super class finalize    | 55   | 13   | 73    | 11    | 0.10  | 4.4     | 1.1      | 541       |

Implementations and speedups are as defined in Sec. 5.1. Queries marked with $^R$ were selected by random sampling.

Table 2: Performance results.

|                                       | M (1/2) | S (1/$3^-$) | H (2/$3^o$) | H' (2/$3^x$) | O (1/$3^o$) | X ($3^o$/$3^x$) | T (1/$3^x$) |
| ------------------------------------- | ------- | ----------- | ----------- | ------------ | ----------- | --------------- | ----------- |
| Geometric means of performance ratios | 2.4x    | 1.2x        | 0.8x        | 5.1x         | 1.9x        | 6.3x            | 12x         |

Table 3: Average performance ratios.

```
for {
  classFile ← classFiles.asSquopt
  method ← classFile.methods
  if method.isAbstract && method.name ==# "equals" &&
  method.descriptor.returnType ==# BooleanType
  parameterTypes ← Let(method.descriptor.parameterTypes)
  if parameterTypes.length ==# 1 && parameterTypes(0) ==#
  classFile.thisClass
} yield (classFile, method)
```

Figure 8: Find covariant `equals` methods.

in Findbugs, we wrote the queries as for-comprehensions in Scala on an in-memory representation created by the Scala toolkit BAT.[4] BAT in particular provides comprehensive support for writing queries against Java bytecode in an idiomatic way. We exemplify an analysis in Fig. 8: It detects all co-variant `equals` methods in a project by iterating over all class files (line 2) and all methods, searching for methods named "equals" that return a boolean value and define a single parameter of the type of the current class.

**Abstractions** In the reference implementations (1), we identified several reusable abstractions as shown in Table 1. The reference implementations of all queries except $17^R$ use exactly one of these abstractions, which encapsulate the main loops of the queries.

---
[4] http://github.com/Delors/BAT

**Indexes** For executing ($3^x$) (SQuOpt with indexes), we have constructed three indexes to speed up navigation over the queried data of queries 1–8: Indexes for method name, exception handlers, and instruction types. We illustrate the implementation of the method-name index in Fig. 9: it produces a collection of all methods and then indexes them using `indexBy`; its argument extracts from an entry the key, that is the method name. We selected which indexes to implement using guidance from SQuOpt itself; during optimizations, SQuOpt reports which indexes it could have applied to the given query. Among those, we tried to select indexes giving a reasonable compromise between construction cost and optimization speedup. We first measured the construction cost of these indexes:

| Index              | Elapsed time (ms)   |
| ------------------ | ------------------- |
| Method name        | 97.99±2.94          |
| Exception handlers | 179.29±3.21         |
| Instruction type   | 4166.49±202.85      |

For our test data, index construction takes less than 200 ms for the first two indexes, which is moderate compared to the time for loading the bytecode in the BAT representation (4755.32 ± 141.66). Building the instruction index took around 4 seconds, which we consider acceptable since this index maps each type of instruction (e.g. `INSTANCEOF`) to a collection of all bytecode instructions of that type.

```
val methodNameIdx: Exp[Map[String, Seq[(ClassFile, Method)]]] =
  (for {
  classFile ← classFiles.asSquopt
  method ← classFile.methods
} yield (classFile, method)).indexBy(entry ⇒ entry._2.name)
```

**Figure 9: A simple index definition**

## 5.3 Measurement Setup

To measure performance, we executed the queries on the preinstalled JDK class library (rt.jar), containing 58M of uncompressed Java bytecode. We also performed a preliminary evaluation by running queries on the much smaller ScalaTest library, getting comparable results that we hence do not discuss. Experiments were run on a 8-core Intel Core i7-2600, 3.40 GHz, with 8 GB of RAM, running Scientific Linux release 6.2. The benchmark code itself is single-threaded, so it uses only one core; however the JVM used also other cores to offload garbage collection. We used the preinstalled OpenJDK Java version 1.7.0_05-icedtea and Scala 2.10.0-M7.

We measure steady-state performance as recommended by Georges et al. [11]. We invoke the JVM $p = 15$ times; at the beginning of each JVM invocation, all the bytecode to analyze is loaded in memory and converted into BAT's representation. In each JVM invocation, we iterate each benchmark until the variations of results becomes low enough. We measure the variations of results through the coefficient of variation (CoV; standard deviation divided by the mean). Thus, we iterate each benchmark until the CoV in the last $k = 10$ iterations drops under the threshold $\theta = 0.1$, or until we complete $q = 50$ iterations. We report the arithmetic mean of these measurements (and also report the usually low standard deviation on our web page).

## 5.4 Results

**Correctness** We machine-checked that all results of all variants of a query agree.

**Modularization overhead** We first observe that performance suffers significantly when using the abstractions we described in Table 1. These abstractions, while natural in the domain and in the setting of a declarative language, are not idiomatic in Java or Scala because, without optimization, they will obviously lead to bad performance. They are still useful abstractions from the point of view of modularity, though—as indicated by Table 1—and as such it would be desirable if one could use them without paying the performance penalty.

**Scala implementations vs. Findbugs** Before actually comparing between the different Scala and SQuOpt implementations, we first ensured that the implementations are comparable to the original Findbugs implementation. A direct comparison between the Findbugs reference implementation and any of our implementations is not possible in a rigorous and fair manner. Findbugs bug detectors are not fully modularized, therefore we cannot reasonably isolate the implementation of the selected queries from support code. Furthermore, the architecture of the implementation has many differences that affect performance: among others, FindBugs also uses multithreading. Moreover, while in our case each query loops over all classes, in FindBugs a single visitor considers each class and invokes all visitors (implemented as listeners) on it.

We measured *startup performance* [11], that is the performance of running the queries only once, to minimize the effect of compiler optimizations. We setup our SQuOpt-based analyses to only perform optimization and run the optimized query. To setup FindBugs, we manually disabled all unrelated bug detectors; we also made the modified Findbugs source code available. The result is that the performance of the Scala implementations of the queries ($3^-$) has performance of the same order of magnitude as the original Findbugs queries – in our tests, the SQuOpt implementation was about twice as fast. However, since the comparison cannot be made fair, we refrained from a more detailed investigation.

**SQuOpt overhead and optimization potential** We present the results of our benchmarks in Table 2. Column names refer to a few of the definitions described above; for readability, we do not present all the ratios previously introduced for each query, but report the raw data. In Table 3, we report the geometric mean [8] of each ratio, computed with the same weight for each query.

We see that, in its current implementation, SQuOpt can cause a overhead S ($1/3^-$) up to 3.4x. On average SQuOpt queries are 1.2x faster. These differences are due to minor implementation details of certain collection operators. For query $18^R$, instead, we have that the the basic SQuOptimplementation is 12.9x faster and are investigating the reason; we suspect this might be related to the use of pattern matching in the original query.

As expected, not all queries benefit from optimizations; out of 19 queries, optimization affords for 15 of them significant speedups ranging from a 1.2x factor to a 12800x factor; 10 queries are faster by a factor of at least 5. Only queries $10^R$, $11^R$ and $12^R$ fail to recover any modularization overhead.

We have analyzed the behavior of a few queries after optimization, to understand why their performance has (or has not) improved.

Optimization makes query $17^R$ slower; we believe this is because optimization replaces filtering by lazy filtering, which is usually faster, but not here. Among queries where indexing succeeds, query 2 has the least speedup. After optimization, it uses the instruction-type index to find all occurrences of invocation opcodes (INVOKESTATIC and INVOKEVIRTUAL); after this step the query looks, among those invocations, for the ones targeting runFinalizersOnExit. Since invocation opcodes are quite frequent, the used index is not very specific, hence it allows for little speedup (9.5x). However no other index applies to this query; moreover, our framework does not maintain any selectivity statistics on indexes to predict these effects. Query $19^R$ benefits from indexing without any specific tuning on our part, because it looks for implementations of finalize with some characteristic, hence the highly selective method-name index applies. After optimization, query 8 becomes simply an index lookup on the index for exception handlers, looking for handlers of IllegalMonitorStateException; it is thus not surprising that its speedup is thus extremely high (12800x). This speedup relies on an index which is specific for this kind of query, and building this index is slower than executing the unoptimized query. On the other hand, building this index is entirely appropriate in a situation where similar queries are common enough. Similar considerations apply to usage of indexing in general, similarly to what happens in databases.

**Optimization overhead** The current implementation of the optimizer is not yet optimized for speed (of the optimization algorithm). For instance, expression trees are traversed and rebuilt completely once for each transformation. However, the optimization overhead is usually not excessive and is $54.8 \pm 85.5$ ms, varying between 3.5 ms and 381.7 ms (mostly depending on the query size).

**Limitations** Although many speedups are encouraging, our optimizer is currently a proof-of-concept and we experienced some limitations:

- In a few cases hand-optimized queries are still faster than what the optimizer can produce. We believe these problems could be addressed by adding further optimizations.

- Our implementation of indexing is currently limited to immutable collections. For mutable collections, indexes must be maintained incrementally. Since indexes are defined as special queries in SQuOpt, incremental index maintenance becomes an instance of incremental maintenance of query results, that is, of incremental view maintenance. We plan to support incremental view maintenance as part of future work; however, indexing in the current form is already useful, as illustrated by our experimental results.

**Summary** We demonstrated on our real-world queries that relying on declarative abstractions in collection queries often causes a significant slowdown. As we have seen, using SQuOpt without optimization, or when no optimizations are possible, usually provides performance comparable to using standard Scala; however, SQuOpt optimizations can in most cases remove the slowdown due to declarative abstractions. Furthermore, relying on indexing allows to achieve even greater speedups while still using a declarative programming style. Some implementation limitations restrict the effectiveness of our optimizer, but since this is a preliminary implementation, we believe our evaluation shows the great potential of optimizing queries to in-memory collections.

# 6. RELATED WORK

This paper builds on prior work on language-integrated queries, query optimization, techniques for DSL embedding, and other works on code querying.

**Language-Integrated Queries** Microsoft's Language-Integrated Query technology (LINQ) [23, 1] is similar to our work in that it also reifies queries on collections to enable analysis and optimization. Such queries can be executed against a variety of backends (such as SQL databases or in-memory objects), and adding new back-ends is supported. Its implementation uses *expression trees*, a compiler-supported implicit conversion between expressions and their reification as a syntax tree. There are various major differences, though. First, the support for expression trees is hard-coded into the compiler. This means that the techniques are not applicable in languages that do not explicitly support expression trees. More importantly, the way expression trees are created in LINQ is generic and fixed. For instance, it is not possible to create different tree nodes for method calls that are relevant to an analysis (such as the map method) than for method calls that are irrelevant for the analysis (such as the toString method). For this reason, expression trees in LINQ cannot be customized to the task at hand and contain too much low-level information. It is well-known that this makes it quite hard to implement programs operating on expression trees [4].

LINQ queries can also not easily be decomposed and modularized. For instance, consider the task of refactoring the filter in the query from x in y where x.z == 1 select x into a function. Defining this function as bool comp(int v) { return v == 1; } would destroy the possibility of analyzing the filter for optimization, since the resulting expression tree would only contain a reference to an opaque function. The function could be declared as returning an expression tree instead, but then this function could not be used in the original query anymore, since the compiler expects an expression of type bool and not an expression tree of type bool. It could only be integrated if the expression tree of the original query is created by hand, without using the built-in support for expression trees.

Although queries against in-memory collections could theoretically also be optimized in LINQ, the standard implementation, LINQ2Objects, performs no optimizations.

A few optimized embedded DSLs allow executing queries or computations on distributed clusters. DryadLINQ [44], based on LINQ, optimizes queries for distributed execution. It inherits LINQ's limitations and thus does not support decomposing queries in different modules. Modularizing queries is supported instead by FlumeJava [3], another library (in Java) for distributed query execution. However, FlumeJava cannot express many optimizations because its representation of expressions is more limited; also, its query language is more cumbersome. Both problems are rooted in Java's limited support for embedded DSLs. Other embedded DSLs support parallel platforms such as GPUs or many-core CPUs, such as Delite [37].

Willis et al. [42, 43] add first-class queries to Java through a source-to-source translator and implement a few selected optimizations, including join order optimization and incremental maintenance of query results. They investigate how well their techniques apply to Java programs, and they suggest that programmers use manual optimizations to avoid expensive constructs like nested loops. While the goal of these works is similar to ours, their implementation as an external source-to-source-translator makes the adoption, extensibility, and composability of their technique difficult.

There have been many approaches for a closer integration of SQL queries into programs, such as HaskellDB [22] (which also inspired LINQ), or Ferry [17] (which moves part of a program execution to a database). In Scala, there are also APIs which integrate SQL queries more closely (e.g., type-checking of queries), such as ScalaQuery[5] and Scala Integrated Query[6]. Based on Ferry, ScalaQL [10] extends Scala with a compiler-plugin to integrate a query language on top of a relational database. The work by Spiewak and Zhao [38] is unrelated to [10] but also called ScalaQL. It is similar to our approach in that it also proposes to reify queries based on for-comprehensions, but it is not clear from the paper how the reification works[7].

---

[5] http://scalaquery.org/
[6] http://code.google.com/p/scala-integrated-query/

**Query Optimization** Query optimization on relational data is a long-standing issue in the database community, but there are also many works on query optimization on objects [7, 14]. Compared to these works, we have only implemented a few simple query optimizations, so there is potential for further improvement of our work by incorporating more advanced optimizations.

**Scala and DSL Embedding** Technically, our implementation of SQuOpt is a deep embedding of a part of the Scala collections API [29]. Deep embeddings were pioneered by Leijen and Meijer [22] and Elliott et al. [5]. The technical details of the embedding are not the main topic of this paper; we are using some of the Scala techniques presented by Rompf and Odersky [36] for using implicits and for adding infix operators to a type. Similar to Rompf and Odersky [36], we also use the Scala compiler on-the-fly. A plausible alternative backend for SQuOpt would have been to use Delite [37], a framework for building highly efficient DSLs in Scala.

We regard the Scala collections API [29] as a shallowly embedded query DSL. Query operators immediately perform collection operations when called, so that it is not possible to optimize queries before execution. In addition to these eager query operators, the Scala collections API also provides *views* to create lazy collections. Views are somewhat similar to SQuOpt in that they reify query operators as data structures and interpret them later. However, views are not used for automatic query optimization, but for explicitly changing the evaluation order of collection processing. Unfortunately, views are not suited as a basis for the implementation of SQuOpt because they only reify the outermost pipeline of collection operators, whereas nested collection operators as well as other Scala code in queries, such as filter predicates or `map` and `flatMap` arguments, are only shallowly embedded. Deep embedding of the whole query is necessary for many optimizations, as discussed in Sec. 3.

**Code Querying** In our evaluation we explore the usage of SQuOpt to express queries on code and re-implement a subset of the Findbugs [20] analyses. There are various other specialized code query languages such as CodeQuest [18] or D-CUBED [41]. Since these are special-purpose query languages that are not embedded into a host language, they are not directly comparable to our approach.

## 7. FUTURE WORK

As part of future work we plan to add support for *incremental view maintenance* [13] to SQuOpt. This would allow, for instance, to update incrementally both indexes and query results.

To make our DSL more convenient to use, it would be useful to use the virtualized pattern matcher of Scala 2.10, when it will be more robust, to add support for pattern matching in our virtualized queries.

Moreover, it would be useful to verify statically that our transformations are type-safe. The optimizer rewrites an expression tree of type `Exp[T]` to another of the same type, but checking this at compile-time is hard because of limitations in the Scala type-checker and its support for GADTs. This is an interesting venue for future work because solving this problem conveniently would make developing transformations safer and thus easier.

## 8. CONCLUSIONS

We have illustrated the tradeoff between performance and modularity for queries on in-memory collections. We have shown that it is possible to design a deep embedding of a version of the collections API which reifies queries and can optimize them at runtime. Writing queries using this framework is, except minor syntactic details, the same as writing queries using the collection library, hence the adoption barrier to using our optimizer is low.

Our evaluation shows that using abstractions in queries introduces a significant performance overhead with native Scala code, while SQuOpt, in most cases, makes the overhead much more tolerable or removes it completely. Since our optimizer is a proof-of-concept with many areas where it can be further improved, a more elaborate version of the optimizer is likely to improve the performance even more, especially in those cases where it cannot yet beat the hand-optimized queries.

**Acknowledgements** The authors would like to thank Sebastian Erdweg for the helpful discussions on this project, Katharina Haselhorst for help with the implementation of the code generator, and the anonymous reviewers for their helpful comments on an earlier draft. This work is supported in part by the European Research Council, grant #203099 "ScalPL".

---

[7] We contacted the authors; they were not willing to provide more details or the sources of their approach.

# APPENDIX

In the next appendixes, we discuss how we implement the interface for writing queries. This discussion is also a case study on how well Scala supports our task.

## A. COLLECTIONS AS A CASE STUDY

As discussed in the introduction, to support optimizations we require a deep embedding of the collections DSL. While the basic idea of deep embedding is well known, it is not obvious how to realize deep embedding when considering the following additional goals:

- To support users adopting SQuOpt, a generic SQuOpt query should share the "look and feel" of the ordinary collections API: In particular, query syntax should remain mostly unchanged. In our case, we want to preserve Scala's *for-comprehension*[8] syntax and its notation for anonymous functions.

- Again to support users adopting SQuOpt, a generic SQuOpt query should not only share the syntax of the ordinary collections API; it should also be well-typed if and only if the corresponding ordinary query is well-typed. This is particularly challenging in the Scala collections library due to its deep integration with advanced type-system features, such as higher-kinded generics and implicit objects [29]. For instance, calling `map` on a `List` will return a `List`, and calling `map` on a `Set` will return a `Set`. Hence the object-language representation and the transformations thereof should be as "typed" as possible. This precludes, among others, a first-order representation of object-language variables as strings.

- SQuOpt should be interoperable with ordinary Scala code and Scala collections. For instance, it should be possible to call normal non-reified functions within a SQuOpt query, or mix native Scala queries and SQuOpt queries.

- The performance of SQuOpt queries should be reasonable even without optimizations. A non-optimized SQuOpt query should not be dramatically slower than a native Scala query. Furthermore, it should be possible to create new queries at run time and execute them without excessive overhead. This goal limits the options of applicable interpretation or compilation techniques.

We think that these challenges characterize deep embedding of queries on collections as a *critical* case study [9] for DSL embedding. That is, it is so challenging that embedding techniques successfully used in this case are likely to be successful on a broad range of other DSLs. From the case study, we report the successes and failures of achieving these goals in SQuOpt.

## B. IMPLEMENTATION: EXPRESSING THE INTERFACE IN SCALA

To optimize a query as described in the previous section, SQuOpt needs to reify, optimize and execute queries. Our implementation assigns responsibility for these steps to three main components: A generic library for reification and execution of general Scala expressions, a more specialized library for reification and execution of query operators, and a dedicated query optimizer. Queries need then to be executed through either compilation (already discussed in Sec. 4.3) or interpretation (to discuss in Sec. B.5). We describe the implementation in more detail in the rest of this section. The full implementation is also available online[9].

A core idea of SQuOpt is to reify Scala code as a data structure in memory. A programmer could directly create instances of that data structure, but we also provide a more convenient interface based on advanced Scala features such as implicit conversions and type inference. That interface allows to automatically reify code with a minimum of programmer annotations, as shown in the examples in Sec. 3. Since this is a case study on Scala's support for deep embedding of DSLs, we also describe in this section how Scala supports our task. In particular, we report on techniques we used and issues we faced.

---
[8]Also known as *for expressions* [30, Ch. 23].
[9]http://www.informatik.uni-marburg.de/~pgiarrusso/SQuOpt

## B.1 Representing expression trees

In the previous section, we have seen that expressions that would have type T in a native Scala query are reified and have type Exp[T] in SQuOpt. The generic type Exp[T] is the base for our reification of Scala expression as expression trees, that is, as data structures in memory. We provide a subclass of Exp[T] for every different form of expression we want to reify. For example, in Fig. 2 the expression author.firstName + " " + author.lastName must be reified even though it is not collection-related, for otherwise the optimizer could not see whether author is used. Knowing this is needed for instance to remove variables which are bound but not used. Hence, this expression is reified as StringConcat(StringConcat(AuthorFirstName(author), Const(" ")), AuthorLastName(author)). This example uses the constructors of the following subclasses of Exp[T] to create the expression tree.

```
case class Const[T](t: T) extends Exp[T]

case class StringConcat(str1: Exp[String],
                        str2: Exp[String])
  extends Exp[String]

case class AuthorFirstName(t: Exp[Author])
  extends Exp[String]

case class AuthorLastName(t: Exp[Author])
  extends Exp[String]
```

Expression nodes additionally implement support code for tree traversals to support optimizations, which we omit here.

This representation of expression trees is well-suited for a representation of the structure of expressions in memory and also for pattern matching (which is automatically supported for case classes in Scala), but inconvenient for query writers. In fact, in Fig. 6 and 5, we have seen that SQuOpt provides a much more convenient front-end: The programmer writes almost the usual code for type T and SQuOpt converts it automatically to Exp[T].

## B.2 Lifting first-order expressions

We call the process of converting from T to Exp[T] *lifting*. Here we describe how we lift first-order expressions – Scala expressions that do not contain anonymous function definitions.

To this end, consider again the fragment author.firstName + " " + author.lastName, now in the context of the SQuOpt-enabled query in Fig. 5. It looks like a normal Scala expression, even syntactically unchanged from Fig. 2. However, evaluating that code in the context of Fig. 5 does not concatenate any strings, but creates an expression tree instead. Although the code looks like the same expression, it has a different *type*, Exp[String] instead of String. This difference in the type is caused by the context: The variable author is now bound in a SQuOpt-enabled query and therefore has type Exp[Author] instead of Author. We can still access the firstName field of author, because expression trees of type Exp[T] provide the same interface as values of type T, except that all operations return expressions trees instead of values.

To understand how an expression tree of type Exp[T] can have the same interface as a value of type T, we consider two expression trees str1 and str2 of type Exp[String]. The implementation of lifting differs depending on the kind of expression we want to lift.

**Method calls and operators** In our example, the operator + should be available on Exp[String], but not on Exp[Boolean], because + is available on String but not on Boolean. Furthermore, we want str1 + str2 to have type Exp[String] and to evaluate not to a string concatenation but to a call of StringConcat, that is, to an expression tree which *represents* str1 + str2. This is a somewhat unusual requirement, because usually, the interface of a generic type does not depend on the type parameters.

To provide such operators and to encode expression trees, we use *implicit conversions* in a similar style as Rompf and Odersky [36]. Scala allows to make expressions of a type T implicitly convertible to a different type U. To this end, one must declare an *implicit conversion function* having type T ⇒ U. Calls to such functions will be inserted by the compiler when required to fix a type mismatch between an expression of type T and a context expecting a value of type U. In addition, a method call e.m(args) can trigger the conversion of e to a type where the method m is present[10]. Similarly, an operator usage, as in str1 + str2, can also trigger an implicit conversion: an expression using an operator, like str1 + str2, is desugared to the method call str1.+(str2), which can trigger an implicit conversion from str1 to a type providing the + method. Therefore from now on we do not distinguish between operators and methods.

To provide the method + on Exp[String], we define an implicit conversion from Exp[String] to a new type providing a + method which creates the appropriate expression node.

```
implicit def expToStringOps(t: Exp[String]) =
  new StringOps(t)
class StringOps(t: Exp[String]) {
  def +(that: Exp[String]): Exp[String] =
    StringConcat(t, that)
}
```

This is an example of the well-known Scala *enrich-my-library pattern*[11] [27].

With these declarations in scope, the Scala compiler rewrites str1 + str2 to expToStringOps(str1).+(str2), which evaluates to StringConcat(str1, str2) as desired. Note that the implicit conversion function expToStringOps is not applicable to Exp[Boolean] because it explicitly specifies the receiver of the +-call to have type Exp[String]. In other words, expressions like str1 + str2 are now *lifted* on the level of expression trees in a type-safe way. For brevity, we refer to the defined operator as Exp[String].+.

**Literal values** However, a string concatenation might also include constants, as in str1 + "foo" or "bar" + str1. To lift str1 + "foo", we introduce a lifting for constants which wraps them in a Const node:

```
implicit def pure[T](t: T): Exp[T] = Const(t)
```

The compiler will now rewrite str1 + "foo" to expToStringOps(str1) + pure("foo"). Similarly,

---
[10]For the exact rules, see Odersky et al. [30, Ch. 21] and Odersky [28].
[11]Also known as *pimp-my-library* pattern.

author. firstName + " " + author.lastName is rewritten to (expToStringOps(author.firstName) + pure(" ")) + author.lastName. Note how the implicit conversions cooperate successfullyto lift the expression.

Analogously, it would be convenient if the similar expression "bar" + str1 would be rewritten to expToStringOps(pure("bar")) + str1, but this is not the case, because implicit coercions are not chained automatically in Scala. Instead, we have to manually chain existing implicit conversions into a new one:

```
implicit def toStringOps(t: String) = expToStringOps(pure(t))
```

so that "bar" + str1 is rewritten to toStringOps("bar") + str1.

**User-defined methods** Calls of user-defined methods like author.firstName are lifted the same way as calls to built-in methods such as string concatenation shown earlier. For the running example, the following definitions are necessary to lift the methods from Author to Exp[Author].

```
package schema.squopt

implicit def expToAuthorOps(t: Exp[Author]) =
  new AuthorOps(t)
implicit def toAuthorOps(t: Author) =
  expToAuthorOps(pure(t))
class AuthorOps(t: Exp[Author]) {
  def firstName: Exp[String] = AuthorFirstName(t)
  def lastName: Exp[String] = AuthorLastName(t)
}
```

Author is not part of SQuOpt or the standard Scala library but an application-specific class, hence it is not possible to pre-define the necessary lifting code as part of SQuOpt. Instead, the application programmer needs to provide this code to connect SQuOpt to his application. To support the application programmer with this tedious task, we provide a code generator which discovers the interface of a class through reflection on its compiled version and generates the boilerplate code such as the one above for Author. It also generates the application-specific expression tree types such as AuthorFirstName as shown in Sec. B.1. In general, query writers need to generate and import the boilerplate lifting code for all application-specific types they want to use in a SQuOpt query.

If desired, we can exclude some methods to *restrict* the language supported in our deeply embedded programs. For instance SQuOpt requires the user to write side-effect-free queries, hence we do not lift methods which perform side effects.

Using similar techniques, we can also lift existing functions and implicit conversions.

**Tuples and other generic constructors** The techniques presented above for the lifting of method calls rely on overloading the name of the method with a signature that involves Exp. Implicit resolution (for method calls) will then choose our lifted version of the function or method to satisfy the typing requirements of the context or arguments of the call. Unfortunately, this technique does not work for tuple constructors, which, in Scala, are not resolved like ordinary calls. Instead, support for tuple types is hard-wired into the language, and tuples are always created by the predefined tuple constructors.

For example, the expression (str1, str2) will always call Scala's built-in Tuple2 constructor and correspondingly have type (Exp[String], Exp[String]). We would prefer that it calls a lifting function and produces an expression tree of type Exp[(String, String)] instead.

Even though we cannot intercept the call to Tuple2, we can add an implicit conversion to be called *after* the tuple is constructed.

```
implicit def tuple2ToTuple2Exp[A1, A2]
    (tuple: (Exp[A1], Exp[A2])): LiftTuple2[A1, A2] =
  LiftTuple2[A1, A2](tuple._1, tuple._2)
case class LiftTuple2[A1, A2](t1: Exp[A1], t2: Exp[A2]) extends
  Exp[(A1, A2)]
```

We generate such conversions for different arities with a code generator. These conversions will be used only when the context requires an expression tree. Note that this technique is only applicable because tuples are generic and support arbitrary components, including expression trees.

In fact, we use the same technique also for other generic constructors to avoid problems associated with shadowing of constructor functions. For example, an implicit conversion is used to lift Seq[Exp[T]] to Exp[Seq[T]]: code like Seq(str1, str2) first constructs a sequence of expression trees and then wraps the result with an expression node that describes a sequence.

**Subtyping** So far, we have seen that for each first-order method m operating on instances of T, we can create a corresponding method which operates on Exp[T]. If the method accepts parameters having types A1, ... , An and has return type R, the corresponding lifted method will accept parameters having types Exp[A1], ... , Exp[An] and return type Exp[R]. However, Exp[T] also needs to support all methods that T inherits from its super-type S. To ensure this, we declare the type constructor Exp to be *covariant* in its type parameter, so that Exp[T] correctly inherits the liftings from Exp[S]. This works even with the enrich-my-library pattern because implicit resolution respects subtyping in an appropriate way.

**Limitations of Lifting** Lifting methods of Any or AnyRef (Scala types at the root of the inheritance hierarchy) is not possible with this technique: Exp[T] inherits such methods and makes them directly available, hence the compiler will not insert an implicit conversion. Therefore, it is not possible to lift expressions such as x == y; rather, we have to rely on developer discipline to use ==# and !=# instead of == and !=.

An expression like "foo" + "bar" + str1 is converted to toStringOps("foo" + "bar") + str1; hence, part of the expression is evaluated before being reified. This is harmless here since we want "foo" + "bar" to be evaluated at compile-time, that is constant-folded, but in other cases it is preferable to prevent the constant folding. We will see later examples where queries on collections are evaluated before reification, defeating the purpose of our framework, and how we work around those.

### B.3 Lifting higher-order expressions

We have shown how to lift first-order expressions; however, the interface of collections also uses higher-order methods, that is, methods that accept functions as parameters, and we need to lift them as well to reify the complete collection DSEL. For instance, the map method applies a function to each element of a collection. In this subsection, we describe how we reify such expressions of function type.

**Higher-order abstract syntax** To represent functions, we have to represent the bound variables in the function bodies. For example, a reification of str ⇒ str + "!" needs to reify the variable str of type String in the body of the anonymous function. This reification should retain the information where str is bound. We achieve this by representing bound variables using *higher-order abstract syntax* (HOAS) [34], that is, we represent them by variables bound at the meta level. To continue the example, the above function is reified as (str: Exp[String]) ⇒ str + "!". Note how the type of str in the body of this version is Exp[String], because str is a reified variable now. Correspondingly, the expression str + "!" is lifted as described in the previous subsection.

With all operations in the function body automatically lifted, the only remaining syntactic difference between normal and lifted functions is the type annotation for the function's parameter. Fortunately, Scala's *local type inference* can usually deduce the argument type from the context, for example, from the signature of the map operation being called. Type inference plays a dual role here: First, it allows the query writer to leave out the annotation, and second, it triggers lifting in the function body by requesting a lifted function instead of a normal function. This is how in Fig. 5, a single call to asSquopt triggers lifting of the overall query.

Note that reified functions have type Exp[A] ⇒ Exp[B] instead of the more regular Exp[A ⇒ B]. We chose the former over the latter to support Scala's syntax for anonymous functions and for-comprehensions which is hard-coded to produce or consume instances of the pre-defined A ⇒ B type. We have to reflect this irregularity in the lifting of methods and functions by treating the types of higher-order arguments accordingly.

**User-defined methods, revised** We can now extend the lifting of signatures for methods or functions from the previous subsection to the general case, that is, the case of higher-order functions. We lift a method or function with signature

**def** m[A1, ..., An](a1: T1, ..., an: Tn): R

to a method or function with the following signature.

**def** m[A1, ..., An](a1: *Lift*⟦T1⟧, ..., an: *Lift*⟦Tn⟧): *Lift*⟦R⟧

As before, the definition of the lifted method or function will return an expression node representing the call. If the original was a function, the lifted version is also defined as a function. If the original was a method on type T, then the lifted version is enriched onto T.

The type transformation *Lift* converts the argument and return types of the method or function to be lifted. For most types, *Lift* just wraps the type in the Exp type constructor, but function types are treated specially: *Lift* recursively descends into function types to convert their arguments separately. Overall, *Lift* behaves as follows.

*Lift*⟦(A1, ..., An) ⇒ R⟧ =
 (*Lift*⟦A1⟧, ..., *Lift*⟦An⟧) ⇒ *Lift*⟦R⟧
*Lift*⟦A⟧ = Exp[A]

We can use this extended definition of method lifting to implement map lifting for Lists, that is, a method with signature Exp[List[T]].map(Exp[T] ⇒ Exp[U]):

```
implicit def expToListOps[T](coll: Exp[List[T]]) =
  new ListOps(coll)
implicit def toListOps[T](coll: List[T]) =
  expToListOps(coll)
class ListOps(coll: Exp[List[T]]) {
```

```
val records = books.
  withFilter(book ⇒ book.publisher == "Pearson Education").
  flatMap(book ⇒ book.authors.
  map(author ⇒ BookData(book.title,
    author.firstName + " " + author.lastName,
    book.authors.size - 1)))
```

Figure 10: Desugaring of code in Fig. 2.

```
  def map[U](f: Exp[T] ⇒ Exp[U]) =
    ListMapNode(coll, Fun(f))
}
case class ListMapNode[T, U](coll: Exp[List[T]], mapping: Exp[T
  ⇒ U]) extends Exp[List[U]]
```

Note how map's parameter f has type Exp[T] ⇒ Exp[U] as necessary to enable Scala's syntax for anonymous functions and automatic lifting of the function body. This implementation would work for queries on lists, but does not support other collection types or queries where the collection type changes. We show in Sec. B.4 how SQUOPT integrates with such advanced features of the Scala Collection DSEL.

## B.4 Lifting collections

In this subsection, we first explain how for-comprehensions are desugared to calls to library functions, allowing an external library to give them a different meaning. We summarize afterwards needed information about the subset of the Scala collection DSEL that we reify. Then we present how we perform this reification. We finally present the reification of the running example (Fig. 5).

**For-comprehensions** As we have seen, an idiomatic encoding in Scala of queries on collections are for-comprehensions. Although Scala is an impure functional language and supports side-effectful for-comprehensions, only pure queries are supported in our framework, because this enables or simplifies many domain-specific analyses. Hence we will restrict our attention to pure queries.

The Scala compiler desugars for-comprehensions into calls to three collection methods, map, flatMap and withFilter, which we explain shortly; the query in Fig. 2 is desugared to the code in Fig. 10.

The compiler performs type inference and type checking on a for-comprehension only *after* desugaring it; this affords some freedom for the types of map, flatMap and withFilter methods.

**The Scala Collection DSEL** A Scala collection containing elements of type T implements the trait Traversable[T]. On an expression coll of type Traversable[T] one can invoke methods declared (in first approximation) as follows:

```
def map[U](f: T ⇒ U): Traversable[U]
def flatMap[U](f: T ⇒ Traversable[U]): Traversable[U]
def withFilter[U](p: T ⇒ Boolean): Traversable[T].
```

For a Scala collection coll, the expression coll.map(f) applies f to each element of coll, collects the results in a new collection and returns it; coll.flatMap(f) applies f to each element of coll, concatenates the results in a new collection and returns it; coll.withFilter(p) produces a collection containing the elements of coll which satisfy the predicate p.

However, Scala supports many different collection types, and this complicates the actual types of these methods. Each collection can further implement sub-traits like `Seq[T] <: Traversable[T]` (for sequences), `Set[T] <: Traversable[T]` (for sets) and `Map[K, V] <: Traversable[(K, V)]` (for dictionaries); for each such trait, different implementations are provided.

One consequence of this syntactic desugaring is that a single for-comprehension can operate over different collection types. The type of the result depends essentially on the type of the root collection, that is `books` in the example above. The example above can hence be altered to produce a sequence rather than a set by simply converting the root collection to another type:

```
val recordsSeq = for {
  book ← books.toSeq
  if book.publisher == "Pearson Education"
  author ← book.authors
} yield BookData(book.title,
    author.firstName + " " + author.lastName,
    book.authors.size - 1)
```

**Precise static typing** The Scala collections DSEL achieves precise static typing while avoiding code duplication [29]. Precise static typing is necessary because the return type of a query operation can depend on subtle details of the base collection and query arguments. To return the most specific static type, the Scala collection DSL defines a type-level relation between the source collection type, the element type for the transformed collection, and the type for the resulting collection. The relation is encoded through the concept pattern [31], i.e., through a type-class-style trait `CanBuildFrom[From, Elem, To]`, and elements of the relation are expressed as implicit instances.

For example, a finite map can be treated as a set of pairs so that mapping a function from pairs to pairs over it produces another finite map. This behavior is encoded in an instance of type `CanBuildFrom[Map[K, V], (K1, V1), Map[K1, V1]]`. The `Map[K, V]` is the type of the base collection, `(K1, V1)` is the return type of the function, and `Map[K1, V1]` is the return type of the map operation.

It is also possible to map some other function over the finite map, but the result will be a general collection instead of a finite map. This behavior is described by an instance of type `CanBuildFrom[Traversable[T], U, Traversable[U]]`. Note that this instance is also applicable to finite maps, because `Map` is a subclass of `Traversable`. Together, these two instances describe how to compute the return type of mapping over a finite map.

**Code reuse** Even though these two use cases for mapping over a finite map have different return types, they are implemented as a single method that uses its implicit `CanBuildFrom` parameter to compute both the static type and the dynamic representation of its result. So the Scala Collections DSEL provides precise typing without code duplication. In our deep embedding, we want to preserve this property.

`CanBuildFrom` is used in the implementation of `map`, `flatMap` and `withFilter`. To further increase reuse, the implementations are provided in a helper trait `TraversableLike[T, Repr]`, with the following signatures:

```
def map[U](f: T ⇒ U)
  (implicit cbf: CanBuildFrom[Repr, U, That]): That
def flatMap[U](f: T ⇒ Traversable[U])
  (implicit cbf: CanBuildFrom[Repr, U, That]): That
def withFilter[U](p: T ⇒ Boolean): Repr.
```

The `Repr` type parameter represents the specific type of the receiver of the method call.

```
class TraversableLikeOps[T, Repr <: Traversable[T]
      with TraversableLike[T, Repr]](t: Exp[Repr]) {
  val t: Exp[Repr]
  def withFilter(f: Exp[T] ⇒ Exp[Boolean]): Exp[Repr] =
    Filter(this.t, FuncExp(f))
  def map[U, That <: TraversableLike[U, That]]
        (f: Exp[T] ⇒ Exp[U])
        (implicit c: CanBuildFrom[Repr, U, That]):
        Exp[That] =
    MapNode(this.t, FuncExp(f))
  //... other methods ...
}
//definitions of MapNode, Filter omitted.
```

**Figure 11: Lifting TraversableLike**

**The lifting** The basic idea is to use the enrich-my-library pattern to lift collection methods from `TraversableLike[T, Repr]` to `TraversableLikeOps[T, Repr]`:

```
implicit def expToTraversableLikeOps[T, Repr]
  (v: Exp[TraversableLike[T, Repr]]) =
  new TraversableLikeOps[T, Repr](v)
```

However, given the design of the collection library, every instance of `TraversableLike[T, Repr]` is also an instance of both `Traversable[T]` and `Repr`; to take advantage of this during interpretation and optimization, we need to restrict the type of `expToTraversableLikeOps`, getting the following conversion:[12]

```
implicit def expToTraversableLikeOps
  [T, Repr <: Traversable[T]
         with TraversableLike[T, Repr]]
  (v: Exp[Repr]) =
  new TraversableLikeOps[T, Repr](v)
```

The query operations are defined in class `TraversableLikeOps[T, Repr]`; a few examples are shown in Fig. 11.[13]

Note how the lifted query operations use `CanBuildFrom` to compute the same static return type as the corresponding non-lifted query operation would compute. This reuse of type-level machinery allows SQUOPT to provide the same interface as the Scala Collections DSEL.

**Code reuse, revisited** We already saw how we could reify `List[T].map` through a specific expression node, `ListMapNode`. However, this approach would require generating many variants for different collections with slightly different types; writing an optimizer able to handle all such variations would be unfeasible because of the amount of code duplication required. Instead, by reusing Scala type-level machinery, we obtain a reification which is statically typed and at the same time avoids code duplication in both our lifting and our optimizer, and in general in all possible consumers of our reification, making them feasible to write.

## B.5 Interpretation

After optimization, SQUOPT needs to interpret the optimized expression trees to perform the query. Therefore,

---

[12]Due to type inference bugs, the actual implicit conversion needs to be slightly more complex, to mention T directly in the argument type. We reported the bug at https://issues.scala-lang.org/browse/SI-5298.

[13]Similar liftings are introduced for traits similar to `TraversableLike`, like `SeqLike`, `SetLike`, `MapLike`, and so on.

the trait Exp[T] declares an **def** interpret(): T method, and each expression node overrides it appropriately to implement a mostly standard typed, tagless [2], environment-based interpreter. The interpreter computes a value of type T from an expression tree of type Exp[T]. This design allows query writers to extend the interpreter to handle application-specific operations. In fact, the lifting generator described in Sec. B.3 automatically generates appropriate definitions of interpret for the lifted operations.

For example, the interpretation of string concatenation is simply string concatenation, as shown in the following fragment of the interpreter. Note that type-safety of the interpreter is checked automatically by the Scala compiler when it compiles the fragments.

```
case class StringConcat(str1: Exp[String],
    str2: Exp[String]) extends Exp[String] {
  def interpret() = str1.interpret() + str2.interpret()
}
```

The subset of Scala we reify roughly corresponds to a typed lambda calculus with subtyping and type constructors. It does not include constructs for looping or recursion, so it should be strongly normalizing as long as application programmers do not add expression nodes with non-terminating interpretations. However, query writers can use the host language to create a reified expression of infinite size. This should not be an issue if SQUOPT is used as a drop-in replacement for the Scala Collection DSEL.

During optimization, nodes of the expression tree might get duplicated, and the interpreter could, in principle, observe this sharing and treat the expression tree as a DAG, to avoid recomputing results. Currently, we do not exploit this, unlike during compilation.

## B.6 Optimization

Our optimizer is structured as a pipeline of different transformations on a single intermediate representation, constituted by our expression trees. Each phase of the pipeline, and the pipeline as a whole, produce a new expression having the same type as the original one. Most of our transformations express simple rewrite rules with or without side conditions, which are applied on expression trees from the bottom up and are implemented using Scala's support for pattern matching [6].

Some optimizations, like filter hoisting (which we applied manually to produce the code in Fig. 4), are essentially domain-specific and can improve complexity of a query. To enable such optimizations to trigger, however, one needs often to perform inlining-like transformations and to simplify the result. Inlining-related transformation can for instance produce code like (x, y)._1, which we simplify to x, reducing abstraction overhead but also (more importantly) making syntactically clear that the result does not depend on y, hence might be computed before y is even bound. This simplification extends to user-defined product types; with definitions in Fig. 1 code like BookData(book.title, ...).title is simplified to book.title.

We have implemented thus optimizations of three classes:

- general-purpose simplifications, like inlining, compile-time beta-reduction, constant folding and reassociation on primitive types, and other simplifications[14];

- domain-specific simplifications, whose main goal is still to enable more important optimizations;

- domain-specific optimizations which can change the complexity class of a query, such as filter hoisting, hash-join transformation or indexing.

Among domain-specific simplifications, we implement a few described in the context of the monoid comprehension calculus [15, 14], such as query unnesting and fusion of bulk operators. Query unnesting allows to *unnest* a for comprehension nested inside another, and produce a single for-comprehension. Furthermore, we can fuse different collection operations together: collection operators like map, flatMap and withFilter can be expressed as folds producing new collections which can be combined. Scala for-comprehension are however more general than monoid comprehensions[15], hence to ensure safety of some optimizations we need some additional side conditions[16].

**Manipulating functions** To be able to inspect a HOAS function body funBody: Exp[S] ⇒ Exp[T], like str ⇒ str + "!", we convert it to *first-order* abstract syntax (FOAS), that is to an expression tree of type Exp[T]. To this end, we introduce a representation of variables and a generator of fresh ones; since variable names are auto-generated, they are internally represented simply as integers instead of strings for efficiency.

To convert funBody from HOAS to FOAS we apply it to a fresh variable v of type TypedVar[S], obtaining a first-order representation of the function body, having type Exp[T], and containing occurrences of v.

This transformation is hidden into the constructor Fun, which converts Exp[S] ⇒ Exp[T], a representation of an expression with one free variable, to Exp[S ⇒ T], a representation of a function.

```
case class App[S, T](f: Exp[S ⇒ T], t: Exp[S]) extends Exp[T]
def Fun[-S, +T](f: Exp[S] ⇒ Exp[T]): Exp[S ⇒ T] = {
  val v = Fun.gensym[S]()
  FOASFun(funBody(v), v)
}
case class FOASFun[S, T](val foasBody: Exp[T], v: TypedVar[S])
  extends Exp[S ⇒ T]
implicit def app[S, T](f: Exp[S ⇒ T]): Exp[S] ⇒ Exp[T] =
  arg ⇒ App(f, arg)
```

Conversely, function applications are represented using the constructor App; an implicit conversion allows App to be inserted implicitly. Whenever f can be applied to arg and f is an expression tree, the compiler will convert f(arg) to app(f)(arg), that is App(f, arg).

In our example, Fun(str ⇒ str + "!") produces FOASFun(StringConcat(TypedVar[String](1), Const("!")), TypedVar[String](1)).

Since we auto-generate variable names, it is easiest to implement represent variable occurrences using the Barendregt convention, where bound variables must always be globally unique; we must be careful to perform renaming after beta-reduction to restore this invariant [35, Ch. 6].

---

[14]Beta-reduction and simplification are run in a fixpoint loop [32]. Termination is guaranteed because our language does not admit general recursion.

[15]For instance, a for-comprehension producing a list cannot iterate over a set.

[16]For instance, consider a for-comprehension producing a set and nested inside another producing a list. This comprehension does not correspond to a valid monoid comprehension (see previous footnote), and query unnesting does not apply here: if we unnested the inner comprehension into the outer one, we would not perform duplicate elimination on the inner comprehension, affecting the overall result.

We can now easily implement substitution and beta-reduction and through that, as shown before, enable other optimizations to trigger more easily and speedup queries.

## C. DISCUSSION

In this section we discuss the degree to which SQUOPT fulfilled our original design goals, and the conclusions for host and domain-specific language design.

## C.1 What worked well

Several features of Scala contributed greatly to the success we achieved. With implicit conversions, the lifting can be made mostly transparent. The advanced type system features were quite helpful to make the expression tree representation typed. The fact that for-comprehensions are desugared *before* type inference and type checking was also a prerequisite for automatic lifting. The syntactic expressiveness and uniformity of Scala, in particular the fact that custom types can have the same look-and-feel as primitive types, were also vital to lift expressions on primitive types.

## C.2 Limitations

Despite these positive experiences and our experimental success, our embedding has a few significant limitations.

The first limitation is that we only lift a subset of Scala, and some interesting features are missing. We do not support *statements* in nested blocks in our queries, but this could be implemented reusing techniques from Delite [37]. More importantly for queries, *pattern matching* cannot be supported by deep embedding similar to ours. In contrast to for-comprehension syntax, pattern matching is desugared only *after* type checking [6], which prevents us from lifting pattern matching notation. More specifically, because an extractor [6] cannot return the representation of a result value (say Exp[Boolean]) to later evaluate; it must produce its final result at pattern matching time. There is initial work on "virtualized pattern matching"[17], and we hope to use this feature in future work.

We also experienced problems with operators that cannot be overloaded, such as == or **if-else** and with lifting methods in scala.Any, which forced us to provide alternative syntax for these features in queries. The Scala-virtualized project [25] aims to address these limitations; unfortunately, it advertises no change on the other problems we found, which we subsequently detail.

It would also be desirable if we could enforce the absence of side effects in queries, but since Scala, like most practical programming languages except Haskell, does not track side effects in the type system this does not seem to be possible.

Finally, compared to *lightweight modular staging* [36] (the foundation of Delite) and to polymorphic embedding [19], we have less static checking for some programming errors when writing queries; the recommended way to use Delite is to write a DSEL program in one trait, in terms of the DSEL interface only, and combine it with the implementation in another trait. In polymorphic embedding, the DSEL program is a function of the specific implementation (in this case, semantics). Either approach ensures that the DSL program is *parametric* in the implementation, and hence cannot refer to details of a specific implementation. However, we judged the syntactic overhead for the programmer to be too high to use those techniques – in our implementation we rely on encapsulation and on dynamic checks at query construction time to achieve similar guarantees.

The choice of interpreting expressions turned out to be a significant performance limitation. It could likely be addressed by using Delite and lightweight modular staging instead, but we wanted to experiment with how far we can get *within* the language in a well-typed setting.

## C.3 What did not work so well: Scala type inference

When implementing our library, we often struggled against limitations and bugs in the Scala compiler, especially regarding type inference and its interaction with implicit resolution, and we were often constrained by its limitations. Not only Scala's type inference is not complete, but we learned that its behavior is only specified by the behavior of the current implementation: in many cases where there is a clear desirable solution, type inference fails or finds an incorrect substitution which leads to a type error. Hence we cannot distinguish, in the discussion, the Scala language from its implementation. We regard many of Scala's type inference problems as bugs, and reported them as such when no previous bug report existed, as noted in the rest of this section. Some of them are long-standing issues, others of them were accepted, for other ones we received no feedback yet at the time of this writing, and another one was already closed as WONTFIX, indicating that a fix would be possible but have excessive complexity for the time being.[18].

**Overloading** The code in Fig. 5 uses the lifted BookData constructor. Two definitions of BookData are available, with signatures BookData(String, String, Int) and BookData(Exp[String], Exp[String], Exp[Int]), and it seems like the Scala compiler should be able to choose which one to apply using overload resolution. This however is not supported simply because the two functions are defined in different scopes[19], hence importing BookData(Exp[String], Exp[String], Exp[Int]) shadows locally the original definition.

**Type inference vs implicit resolution** The interaction between type inference and implicit resolution is a hard problem, and Scalac has also many bugs, but the current situation requires further research; for instance, there is not even a specification for the behavior of type inference[20].

As a consequence, to the best of our knowledge some properties of type inference have not been formally established. For instance, a reasonable user expectation is that removing a call to an implicit conversion does not alter the program, if it is the only implicit conversion with the correct type in scope, or if it is more specific than the others [30, Ch. 21]. This is not always correct, because removing the implicit conversion reduces the information available for the type inference algorithm; we observed multiple cases[21] where type inference becomes unable to figure out enough information about the type to trigger implicit conversion.

We also consider significant that Scala 2.8 required making both type inference and implicit resolution more powerful, specifically in order to support the collection library [24, 30, Sec 21.7]; further extensions would be possible and desirable. For instance, if type inference were extended with higher-order unification[22] [33], it would better support a part of our DSL interface (not discussed in this paper) by removing the need for type annotations.

---

[17]http://stackoverflow.com/questions/8533826/what-is-scalas-experimental-virtual-pattern-matcher

[18]https://issues.scala-lang.org/browse/SI-2551
[19]https://issues.scala-lang.org/browse/SI-2551
[20]https://issues.scala-lang.org/browse/SI-5298?focusedCommentId=55971#comment-55971, reported by us.
[21]https://issues.scala-lang.org/browse/SI-5592, reported by us.

**Nested pattern matches for GADTs in Scala** Writing a typed decomposition for `Exp` requires pattern-matching support for generalized algebraic datatypes (GADTs). We found that support for GADTs in Scala is currently insufficient. Emir et al. [6] define the concept of *typecasing*, essentially a form of pattern-matching limited to non-nested patterns, and demonstrate that Scala supports typecasing on GADTs in Scala by demonstrating a typed evaluator; however, typecasing is rather verbose for deep patterns, since one has to nest multiple pattern-matching expressions. When using normal pattern matches, instead, the support for GADT seems much weaker.[23] Hence one has to choose between support for GADT and the convenience of nested pattern matching. A third alternative is to ignore or disable compiler warnings, but we did not consider this option.

**Implicit conversions do not chain** While implicit conversions by default do not chain, it is sometimes convenient to allow chaining selectively. For instance, let us assume a context such that `a: Exp[A]`, `b: Exp[B]` and `c: Exp[C]`. In this context, let us consider again how we lift tuples. We have seen that the expression `(a, b)` has type `(Exp[A], Exp[B])` but can be converted to `Exp[(A, B)]` through an implicit conversion. Let us now consider *nested* tuples, like `((a, b), c)`: it has type `((Exp[A], Exp[B]), Exp[C])`, hence the previous conversion cannot be applied to this expression.

Odersky et al. [30, Ch. 21] describe a pattern which can address this goal. Using this pattern, to lift pairs, we must write an implicit conversion from pairs of elements which can be *implicitly converted* to expressions. Instead of requiring `(Exp[A], Exp[B])`, the implicit conversion should require `(A, B)` with the condition that `A` can be converted to `Exp[A']` and `B` to `Exp[B']`. This conversion solves the problem if applied explicitly, but the compiler refuses to apply it implicitly, again because of type inference issues[24].

Because of this type inference limitations, we failed to provide support for reifying code like `((a, b), c)`[25].

**Error messages for implicit conversions** The enrich-my-library pattern has the declared goal to allow to extend existing libraries *transparently*. However, implementation details shine however through when a user program using this feature contains a type error. When invoking a method would require an implicit conversion which is not applicable, the compiler often just reports that the method is not available. The recommended approach to debugging such errors is to manually add the missing implicit conversion and investigating the type error [30, Ch. 21.8], but this destroys the transparency of the approach when creating or modifying code. We believe this could be solved in principle by research on error reporting: the compiler could automatically insert all implicit conversions enabling the method calls and report corresponding errors, even if at some performance cost.

## C.4 Lessons for language embedders

Various domains, such as the one considered in our case study, allow powerful domain-specific optimizations. Such optimizations often are hard to express in a compositional way, hence they cannot be performed while building the query but must be expressed as global optimizations passes. For those domains, deep embedding is key to allow significant optimizations. On the other hand, deep embedding requires to implement an interpreter or a compiler.

On the one hand, interpretation overhead is significant in Scala, even when using HOAS to take advantage of the metalevel implementation of argument access.

Instead of interpreting a program, one can compile a DSEL program to Scala and load it, as done by Rompf et al. [37]; while we are using this approach, the disadvantage is the compilation delay, especially for Scala whose compilation process is complex and time-consuming. Possible alternatives include generating bytecode directly or combining interpretation and compilations similarly to tiered JIT compilers, where only code which is executed often is compiled. We plan to investigate such alternatives in future work.

---

[22]https://issues.scala-lang.org/browse/SI-2712
[23]Due to bug https://issues.scala-lang.org/browse/SI-5298?focusedCommentId=56840#comment-56840, reported by us.
[24]https://issues.scala-lang.org/browse/SI-5651, reported by us.
[25]One could of course write a specific implicit conversions for *this* case; however, `(a, (b, c))` requires already a different implicit conversion, and there are infinite ways to nest pairs, let alone tuples of bounded arity.